\documentclass{article}
\usepackage{spconf,amsmath,graphicx,multirow}
\usepackage{tikz}
\usetikzlibrary{arrows,positioning,calc,shapes,shapes.geometric}
\tikzset{>=latex}
\usepackage{pgfplots}
\pgfplotsset{compat=1.14}
\usepackage{subfig}
\newlength\figureheight 
\newlength\figurewidth 
\title{Eventness: Object Detection on Spectrograms for Temporal Localization of Audio Events}
\begin{document}
 \name{\begin{tabular}{c}Phuong Pham, Juncheng Li, Joseph Szurley, Samarjit Das\end{tabular}}
 \address{phuongpham@cs.pitt.edu, junchenl@cs.cmu.edu, \{joseph.szurley,samarjit.das\}@us.bosch.com}

%
\maketitle

\begin{abstract}
In this paper, we introduce the concept of \emph{Eventness} for audio event detection, which can, in part, be thought of as an analogue to \emph{Objectness} from computer vision.  The key observation behind the eventness concept is that audio events reveal themselves as 2-dimensional time-frequency patterns with specific textures and geometric structures in spectrograms. These time-frequency patterns can then be viewed analogously to objects occurring in natural images (with the exception that scaling and rotation invariance properties do not apply). With this key observation in mind, we pose the problem of detecting monophonic or polyphonic audio events as an equivalent visual object(s) detection problem under \emph{partial occlusion} and clutter in spectrograms. We adapt a state-of-the-art visual object detection model to evaluate the audio event detection task on publicly available datasets.  The proposed network has comparable results with a state-of-the-art baseline and is more robust on minority events. Provided large-scale datasets, we hope that our proposed conceptual model of eventness will be beneficial to the audio signal processing community towards improving performance of audio event detection.   
\end{abstract}

\section{Introduction}
Recently, objectness-based deep learning networks using region proposals have achieved state-of-the-art performance in natural object detection tasks \cite{girshick2014rich, szegedy2014scalable, renNIPS15fasterrcnn, redmon2016you}. In \cite{girshick2014rich}, a dramatic improvement in object detection was achieved when feeding a convolutional neural network (CNN) with class-agnostic \textit{objectness} region proposals. 
This was later extend to a so-called MultiBox model \cite{szegedy2014scalable}, which integrated the objectness region proposal component into a deep neural network further improving performance.  

One drawback of these methods, however, is the large number of parameters utilized for different features of the region proposals causing a dramatic decrease in processing speed.
Ren et al. \cite{renNIPS15fasterrcnn} overcame this decrease in processing speed by implementing a so-called Faster R-CNN which shares the feature map between the \textit{objectness} region proposal and the object classifier. This unification was shown to not only increase the processing speed but also improve performance.

Audio event detection has recently been investigated from a similar computer vision perspective that exploits the effectiveness of deep CNNs \cite{takahashi2016deep,piczak2015environmental}.  In order to exploit vision inspired CNNs for audio event detection, the audio signal is usually converted to a 2D time-frequency representation, or spectrogram.  By using multiple convolutional layers \cite{piczak2015environmental} or multiple convolutional groups \cite{takahashi2016deep},  these computer vision inspired CNNs have become state-of-the-art in terms of performance for both event detection and classification.  

In this paper, we borrow from the concept of \emph{Objectness} and propose a similar analogue for audio signals termed \emph{Eventness}.  When represented in the time-frequency domain, audio events reveal themselves as 2D patterns in spectrograms, where each event has a specific geometric structure.  These geometric structures then provide information on how the frequencies that comprise the audio event vary with time.  The patterns in the spectrograms can then be thought of synonymously to objects occurring in natural images.  We therefore look to leverage a number of components from objectness-based deep learning and harness them for temporal localization (detection) of audio events from spectrograms\footnote{We note that the same rotation and scale invariance properties of natural images do necessarily translate to the audio domain.}.  

For our proposed \textit{eventness model}, we adapt a state-of-the-art object detection network, namely a Faster R-CNN, for audio event detection.  Audio signals are first converted into spectrograms and a linear intensity mapping is used to separate the spectrogram into 3 distinct channels.  A pre-trained vision based CNN is then used to extract feature maps from the spectrograms, which are then fed into the Faster R-CNN.  It should be noted that while the feature maps produced by the pre-trained vision based CNN aim to detect objects in natural images, they can also be employed in a similar fashion to detect 2D patterns that are present in spectrograms.

To the best of our knowledge, this is the first time audio event detection, or temporal localization, has been approached from a vision-inspired angle, i.e., \emph{objectness} in audio spectrograms.  Compared to other CNN based audio event detection models, ours differs in at least two aspects. First, we propose the regions of events directly instead of inferring these values based on: the concatenation of heuristics via merging  same-class neighbors \cite{plinge2014bag} \cite{cakir2015polyphonic}, performing sequence labeling using HMM and Viterbi algorithms \cite{mesaros2010acoustic} \cite{gemmeke2013exemplar}, or estimating the distance between the current time and the onset and offset of the event \cite{phan2015random}. Second, previous models use small temporal windows, typically between $25ms - 100ms$, \cite{plinge2014bag,cakir2015polyphonic,mesaros2010acoustic,gemmeke2013exemplar,phan2015random} which may not capture complete non-speech events \cite{1599-17}.  Therefore we use large temporals windows on the order of seconds or tens of seconds.   

We evaluate the model on publicly available datasets, namely the UrbanSound8k \cite{salamon2014dataset} and the 2017 Detection and Classification of Acoustic Scenes and Events (DCASE)\cite{mesaros2017dcase} both qualitatively and quantitatively.  In the qualitative analysis we examine the ability of the proposed model to exploit both temporal and spectral content for region proposals, even for overlapping events. 
In the quantitative analysis, we compare both the proposed model and state-of-the-art baseline showing that comparable performance is achieved along with a robustness to infrequent events.
\begin{figure}[!t]
    \centering
    \input{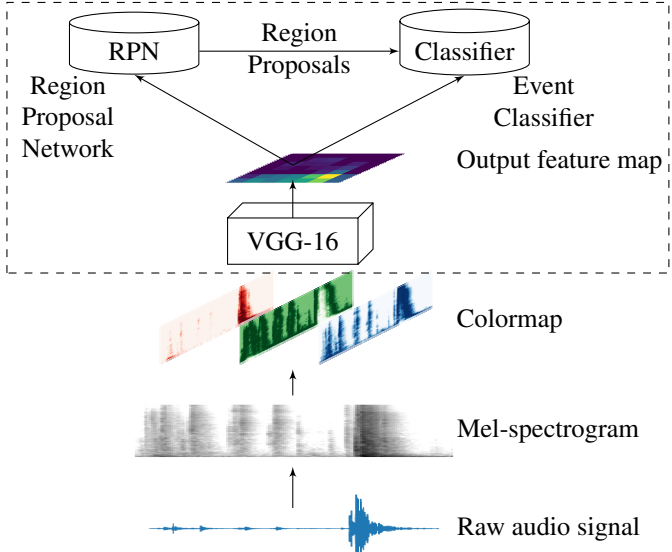}
     \vspace{-10mm}
    \caption{The envisaged eventness model.}
 \label{fig:frcnn}
 \end{figure}
\section{The Eventness model}
Figure \ref{fig:frcnn} shows the envisaged model where audio signals are first converted into log-scaled mel-spectrogams. A feature map is then created by passing the log-scaled mel- spectrogram through a pre-trained CNN based on the VGG-16 model.  This feature map is then fed to a Faster R-CNN consisting of two main components: an RPN and an event classifier, which will be described in Section \ref{subsec:RPN} and \ref{subsec:event_classifier} respectively.  It should be noted that the RPN and event classifier then use the same resulting feature map from the output of the pre-trained VGG-16 CNN to generate the region proposals and classify the audio event. 

\subsection{Spectrogram and feature map generation}
\label{subsec:spectrogram_and_feature_map_generation}
The raw audio signals are first segmented into long time windows of length $T$ and converted to log-scaled mel-spectrograms.  The resulting log-scaled mel-spectrograms are then normalized in the range [0,1] and a linear intensity segment mapping is used to separate the the original spectrogram into 3 distinct channels.  We note that this mapping is the same process used to map greyscale image to a higher dimensional space, i.e., a colormap, which in \cite{dennis2011spectrogram} was shown to outperform single channel inputs of CNN based audio event classifiers.  Other mappings have been proposed based on both the first- and second-order differences of the log-scaled mel-spectrograms, the so-called delta and delta-delta coefficients \cite{takahashi2016deep,piczak2015environmental}.  However, in our  experiments, the previously described linear intensity segment mappings have shown better performance.

The linear intensity segment mapping effectively quantizes the original spectrogram based on the spectral intensities, where strong spectral values will be more prominent in one channel while weak spectral values will be prominent in another.  We hypothesize that the subtle \textit{shapes} introduced by this mapping can be better exploited in the convolutional layers in the network.  The resultant 3-channel spectrogram mapping was then fed into a pre-trained VGG-16 CNN to produce a feature map.  Only the convolutional layers are used from the pre-trained VGG-16 CNN, i.e., we discard the fully connected and classification layers.   
\begin{figure}[!t]
\centering
\includegraphics[width=\columnwidth]{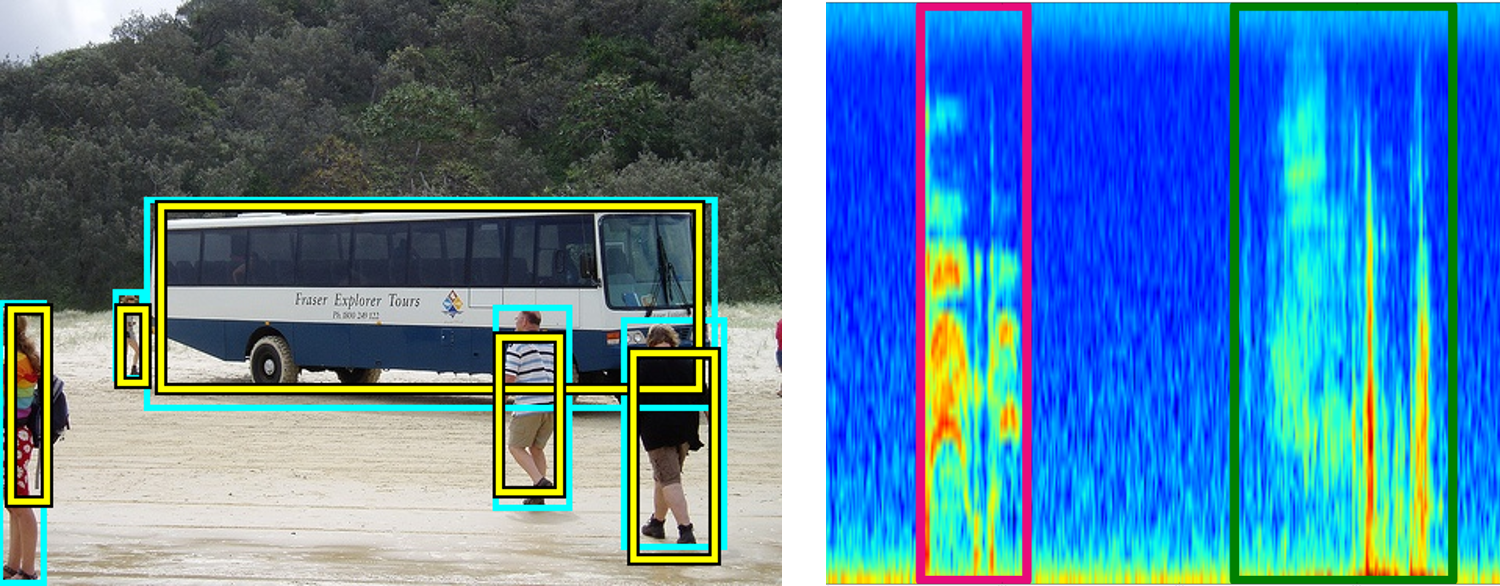}
\caption{\textit{Objectness} proposals in an natural image (left) and audio \textit{eventness} proposals in a spectrogram image (right).}
\label{fig:spectrogram}
\end{figure}
\subsection{Regional Proposal Network (RPN)}
\label{subsec:RPN}
The RPN in the proposed model uses \emph{anchors} to generate multiple region proposals based on the output feature map of the pre-trained VGG-16 CNN. In particular, at each location of the feature map, the RPN will generate multiple region proposals at different scales and aspect ratios. The intuition behind using this anchor approach is that an estimate of a complete object can be inferred even under partial observation or occlusion, e.g., in the left image of Figure \ref{fig:spectrogram} even though the bus is occluded by pedestrians, the RPN still generates a full proposal over the entire bus.  This can then directly be translated to overlapping audio events in both the time and frequency domain.  Furthermore, the anchors from the RPN will generate eventness proposals using both the temporal and spectral content as shown in the right image of Figure \ref{fig:spectrogram}.  
\subsection{Event Classifier}
\label{subsec:event_classifier}
The event classifier employs a Region of Interest (RoI) pooling layer \cite{girshickICCV15fastrcnn} to extract and normalize region proposals from the RPN.  The RoI pooling layer performs a similar operation to a conventional max-pooling layer,  except that it can take inputs of non-uniform sizes to obtain fixed-size feature maps.  

By focusing only on the region proposals provided by the RPN, the classifier can ignore other noisy (unimportant) areas of the feature map. Besides classifying the audio event, this component also refines the region bounding box. The allows for a more accurate bounding box which is derived from a \emph{larger} view of the event instead of only one location in the feature map. The event classifier then uses a final softmax layer to predict the audio event.   
\section{Experiment and results}
\subsection{Datasets and Performance Metrics}
\label{subsec:input_features}
From a given audio file, large audio segments ($T = 15s$ from the UrbanSound8K dataset and $T = 10s$ from the DCASE 2017 dataset) were first extracted. 128-band log-scaled mel-spectrograms were generated from the audio segments with a window size of 2048 and hop length is 1024.  Feature maps were then generated as described in Section \ref{subsec:spectrogram_and_feature_map_generation} and fed to the the RPN and classifier.  The RPN then used 9 anchors at each location on the feature map to generate multiple eventness proposals.    

The performance metrics are used on either a segment based or event based level \cite{mesaros2017dcase}.  The segment-based metrics, $F_{1,SB}$ score and error rate denoted as $ER_{SB}$, are used to compare the predicted events to the ground truth labels of segments that are one second long.  The same metrics are applied for event-based metrics, $F_{1,EB}$ score and error rate denoted as $ER_{EB}$, which compare the amount of overlap between the predicted event and the ground truth labels.  
The definitions of the error rates are as follow:
\begin{align*}
    ER_{SB}&=\frac{\max(N_{ref}, N_{sys}) - TP}{N_{ref}}\\
    ER_{EB}&=\frac{FN + FP}{N_{ref}}
\end{align*}
where $N_{ref}$ is total number of predicted events, $N_{sys}$ is total number ground truths events, and $TP$, $FP$, and $FN$ are the number of true positives, false positive, and false negatives respectively.
  
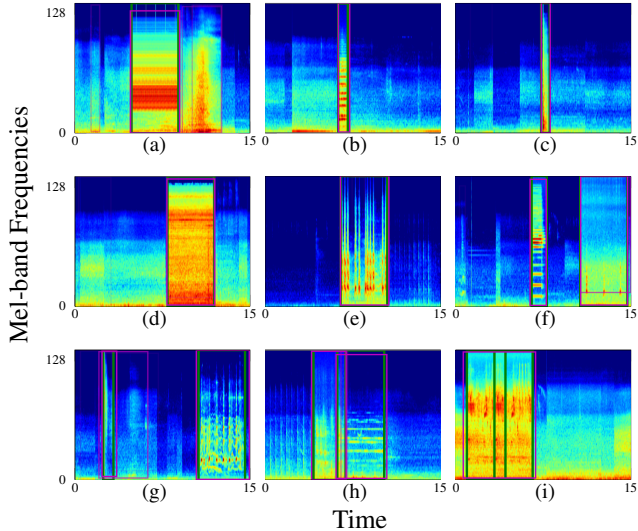
\begin{figure}[!t]
\begin{tikzpicture}[overlay,remember picture]
    \node[rotate=90]
    at (0,-3.8)
    {Mel-band Frequencies};
  \end{tikzpicture}
 
\begin{tikzpicture}[overlay,remember picture]
    \node[]
    at (4.5,-7.3)
    {Time};
  \end{tikzpicture} 
 
\centering
   \setlength\figureheight{.2\columnwidth}
   \setlength\figurewidth{.27\columnwidth}
\subfloat[][]
{
	\begin{tikzpicture}
    \begin{axis} [%
        width=\figurewidth,
		height=\figureheight,
        scale only axis, 
        enlargelimits=false,
        clip=false,xticklabel style = {font=\tiny,yshift=0.5ex},yticklabel style = {font=\tiny},xticklabels={0,0,,,15},yticklabels={0,0,,128,128}]
        \addplot graphics [
            xmin=0,
            xmax=15,
            ymin=0,
            ymax=108,
        ] {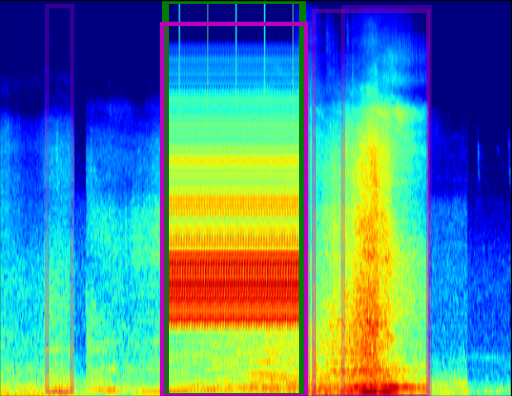};
    \end{axis}

\end{tikzpicture}
	\label{fig:a}
}
\hspace{-5mm}\subfloat[]
{
	\begin{tikzpicture}
    \begin{axis} [%
        width=\figurewidth,
		height=\figureheight,
        scale only axis, 
        enlargelimits=false,
        clip=false,xticklabel style = {font=\tiny,yshift=0.5ex},yticklabel style = {font=\tiny},yticklabels={,,},xticklabels={0,0,,,15}]
        \addplot graphics [
            xmin=0,
            xmax=15,
            ymin=0,
            ymax=128,
        ] {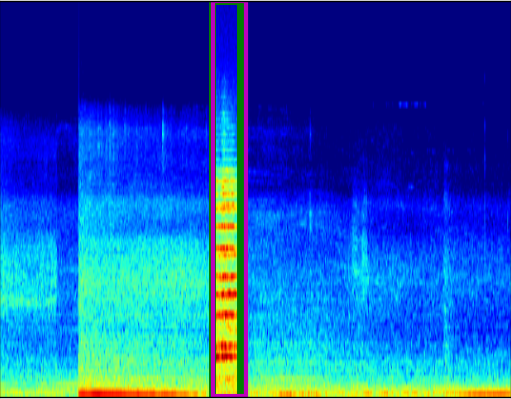};
    \end{axis}

\end{tikzpicture}
	\label{fig:b}
}
\hspace{-5mm}\subfloat[]
{
	\begin{tikzpicture}
    \begin{axis} [%
        width=\figurewidth,
		height=\figureheight,
        scale only axis, 
        enlargelimits=false,
        clip=false,xticklabel style = {font=\tiny,yshift=0.5ex},yticklabel style = {font=\tiny},yticklabels={,,},xticklabels={0,0,,,15}]
        \addplot graphics [
            xmin=0,
            xmax=15,
            ymin=0,
            ymax=128,
        ] {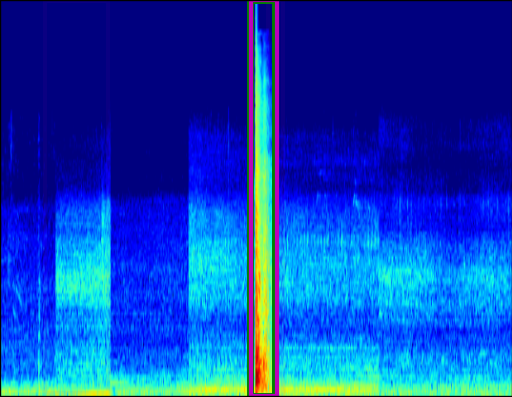};
    \end{axis}

\end{tikzpicture}
	\label{fig:c}
}\\\vspace{-2mm}
\subfloat[]
{
	\begin{tikzpicture}
    \begin{axis} [%
        width=\figurewidth,
		height=\figureheight,
        scale only axis, 
        enlargelimits=false,
        clip=false,xticklabel style = {font=\tiny,yshift=0.5ex},yticklabel style = {font=\tiny},xticklabels={0,0,,,15},yticklabels={0,0,,128}]
        \addplot graphics [
            xmin=0,
            xmax=15,
            ymin=0,
            ymax=108,
        ] {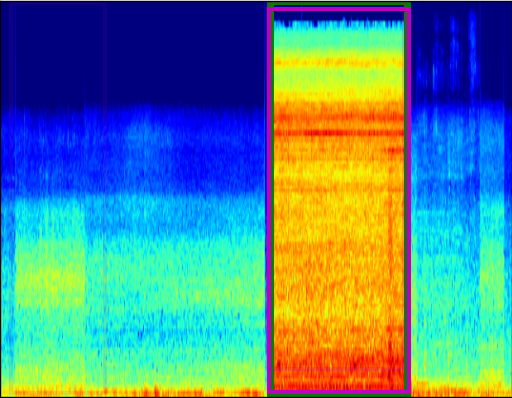};
    \end{axis}

\end{tikzpicture}
	\label{fig:d}
}
\hspace{-5mm}\subfloat[]
{
	\begin{tikzpicture}
    \begin{axis} [%
        width=\figurewidth,
		height=\figureheight,
        scale only axis, 
        enlargelimits=false,
        clip=false,xticklabel style = {font=\tiny,yshift=0.5ex},yticklabel style = {font=\tiny},yticklabels={,,},xticklabels={0,0,,,15}]
        \addplot graphics [
            xmin=0,
            xmax=15,
            ymin=0,
            ymax=128,
        ] {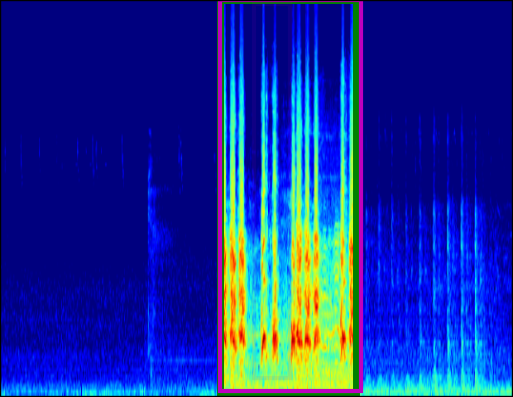};
    \end{axis}

\end{tikzpicture}
	\label{fig:e}
}
\hspace{-5mm}\subfloat[]
{
	\begin{tikzpicture}
    \begin{axis} [%
        width=\figurewidth,
		height=\figureheight,
        scale only axis, 
        enlargelimits=false,
        clip=false,xticklabel style = {font=\tiny,yshift=0.5ex},yticklabel style = {font=\tiny},yticklabels={,,},xticklabels={0,0,,,15}]
        \addplot graphics [
            xmin=0,
            xmax=15,
            ymin=0,
            ymax=128,
        ] {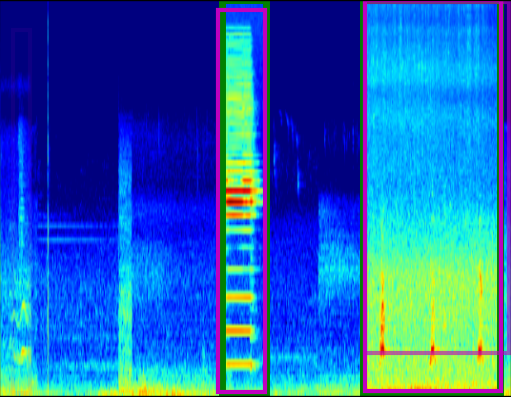};
    \end{axis}

\end{tikzpicture}
	\label{fig:f}
}\\\vspace{-2mm}
\subfloat[]
{
	\begin{tikzpicture}
    \begin{axis} [%
        width=\figurewidth,
		height=\figureheight,
        scale only axis, 
        enlargelimits=false,
        clip=false,xticklabel style = {font=\tiny,yshift=0.5ex},yticklabel style = {font=\tiny},xticklabels={0,0,,,15},yticklabels={0,0,,128}]
        \addplot graphics [
            xmin=0,
            xmax=15,
            ymin=0,
            ymax=108,
        ] {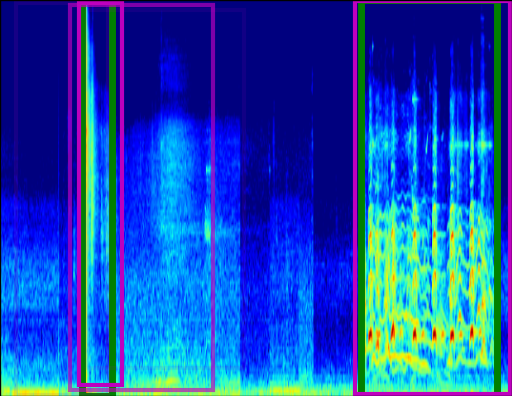};
    \end{axis}

\end{tikzpicture}
	\label{fig:g}
}
\hspace{-5mm}\subfloat[]
{
	\begin{tikzpicture}
    \begin{axis} [%
        width=\figurewidth,
		height=\figureheight,
        scale only axis, 
        enlargelimits=false,
        clip=false,xticklabel style = {font=\tiny,yshift=0.5ex},yticklabel style = {font=\tiny},yticklabels={,,},xticklabels={0,0,,,15}]
        \addplot graphics [
            xmin=0,
            xmax=15,
            ymin=0,
            ymax=128,
        ] {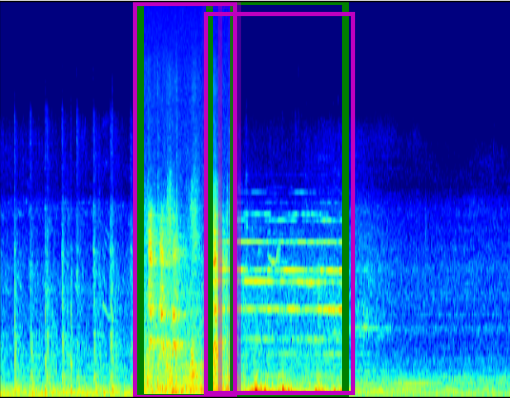};
    \end{axis}

\end{tikzpicture}
	\label{fig:h}
}
\hspace{-5mm}\subfloat[]
{
	\begin{tikzpicture}
    \begin{axis} [%
        width=\figurewidth,
		height=\figureheight,
        scale only axis, 
        enlargelimits=false,
        clip=false,xticklabel style = {font=\tiny,yshift=0.5ex},yticklabel style = {font=\tiny},yticklabels={,,},xticklabels={0,0,,,15}]
        \addplot graphics [
            xmin=0,
            xmax=15,
            ymin=0,
            ymax=128,
        ] {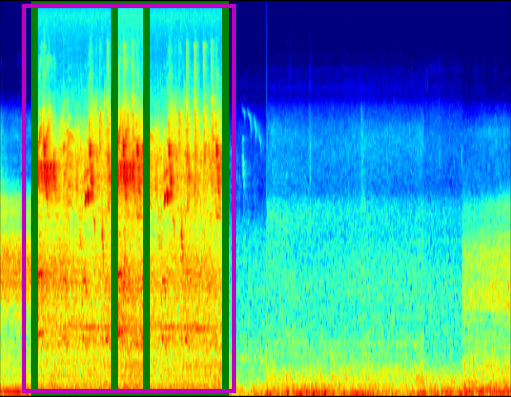};
    \end{axis}

\end{tikzpicture}
	\label{fig:i}
}
\caption{Region proposals of several audio events for  (a) siren, (b) car horn , (c) gun shot, (d) jackhammer, (e) dog bark, (f) car horn with dog bark, (g) gun shot with dog bark, (h) dog bark with siren, (i) dog bark with dog bark.}
\label{fig:qualitative}
\end{figure}
\subsection{Qualitative Evaluation}
\label{subsec:qual_eval}
We perform a qualitative analysis of the proposed model on the UrbanSound8K dataset which consists of more than 8000 audio clips comprised of 10 classes.  We select 5 target classes having good performance in a pilot experiment out of 10 classes, specifically: car horn, dog bark, gunshot, siren, and jackhammer. We then randomly embed one or two events from these classes with background noise from the DCASE 2016 (task 3) \cite{mesaros2016tut} dataset which is meant to mimic multisource conditions.  This produced, in total, 5000 training clips and 3000 testing clips, with 30\% being polyphonic.

Figure \ref{fig:qualitative} shows 5 monophonic spectrograms, i.e. siren (a), car horn (b), gun shot (c), jackhammer (d), and dog bark (e), and 4 polyphonic spectrograms, i.e. car horn-dog bark (f), gun shot-dog bark (g), dog bark-siren (h), and dog bark-dog bark (i).  The Figure also has an overlay of the region proposals where the areas surrounded by the green boxes are the ground truth labels and the areas surrounded by red boxes are the predicted labels.  The darker the shade of red indicates a higher confidence level of the prediction.  

In general, the region proposals are able to capture the audio events with a very fine temporal resolution.  In Figure \ref{fig:qualitative} (a), besides correctly identifying the siren audio event, the eventness model surprisingly detected another audio event that was erroneously omitted from the ground truth labels.  Furthermore in Figure \ref{fig:qualitative} (g) the eventness model detects an audio event that is corrupted by noise and again erroneously omitted from the ground truth labels.  

In Figure \ref{fig:qualitative} (h), the eventness model detects events with different temporal lengths, i.e., the dog bark event is shorter than the siren event and can differentiate between partially overlapping events. Furthermore, the proposed model also provides region proposals based on the spectral content, as noticed by their sizes in along the frequency axis, i.e., the vertical size of the bounding box is different for each unique event.  

In Figure \ref{fig:qualitative} (i), the same audio event occurs with some overlap.  Remarkably, the eventness model extrapolates this as a single event.  This is a direct result of using larger temporal windows as the eventness model has a larger global view when compared to other work using smaller temporal windows.
\begin{table}[t!]
\caption{Performance of the eventness model on the UrbanSound8K dataset.}
\label{tab:qualitativePerf}
\vspace{3mm}
\centering
\resizebox{\columnwidth}{!}{%
\begin{tabular}{|l|c|c|c|c|c|c|}
  \hline
  Metric&Car horn&Dog bark&Gun shot&Jackhammer&Siren&Overall\\
  \hline
  $F_{1,EB}$&0.86&0.61&0.79&0.75&0.55&0.71\\
  \hline
  $ER_{EB}$&0.27&0.81&0.40&0.52&0.70&0.54\\
  \hline
\end{tabular}
}
\end{table}
\subsection{Quantitative Evaluation}
\label{subsec:quan_eval}
We quantitatively evaluate the eventness model on the UrbanSound8k dataset as described in Section \ref{subsec:qual_eval} and the DCASE 2017 task 3 dataset. The DCASE dataset is pre-divided into 4 folds, where we select the first fold for testing and synthesize 10s audio clips from the remaining folds which contain 11,260 clips in total)\footnote{Even though evaluation was performed on a single fold, we expect that performing 4 fold cross validation result would not significantly impact the accuracy.}. Synthesized clips are generated by randomly assigning one or two annotated events with the background noise (audio portion having no ground truth annotations) from the DCASE 2017. Even when we select an event to synthesize, the selected event is usually overlapped with some other events due to the natural recording environment. Audio clips in the test set are kept unchanged.

Table \ref{tab:qualitativePerf} shows both the $F_{1,EB}$ and $ER_{EB}$ for the UrbanSound8k dataset.  We see that the eventness model has a high F1 score, meaning it has both good recall and precision and a low error rate.  

Table \ref{tab:detectionPerf} shows the performance of the proposed \textit{eventness} model and the baseline model provided by DCASE 2017 which utilizes a 2 layer neural network with small temporal windows, $T=40ms$.  It can be seen that the eventness model outperforms the baseline in event-based metrics but not in segment-based metrics.  This is due to the eventness model focusing on larger temporal segments which capture the entire event whereas the baseline leverages smaller temporal segments which are better suited for segment-based tasks.  

Interestingly, the eventness model exhibits more robustness to infrequent events when compared to the baseline model.  The third event class, children  is the least frequent classes in the dataset and the error rate using the eventness model has superior performance, although both models have similar F1 scores.  
\begin{table}[t!]
\caption{Performance on DCASE 2017. The reported numbers are in the following format $ER$ and $(F_1)$}
\label{tab:detectionPerf}
\vspace{3mm}
\centering
\resizebox{\columnwidth}{!}{%
\begin{tabular}{|l|c|c|c|c|}
  \hline
  \multirow{2}{*}{Event class}&\multicolumn{2}{|c|}{Segment-based} & \multicolumn{2}{|c|}{Event-based}\\
  \cline{2-5}
  &Baseline&Eventness&Baseline&Eventness\\
  \hline
  Brakes squeaking&1(0)&0.97(0.29)&1(0)&1.75(0)\\
  \hline
Car&0.86(0.69)&1.12(0.61)&2.66(0.05)&1.55(0.04)\\
  \hline
Children&5.45(0)&1.86(0)&13(0)&2.18(0)\\
  \hline
Large vehicle&1.15(0.31)&1.36(0.28)&5.92(0.03)&3.31(0)\\
  \hline
People speaking&1.34(0.03)&1.01(0.10)&3.47(0)&1.2(0.08)\\
  \hline
People walking&1.07(0.28)&1.08(0.25)&3.88(0.04)&1.81(0.03)\\
  \hline
Overall&0.95(0.45)&1.02(0.42)&3.53(0.03)&1.78(0.03)\\
  \hline
\end{tabular}
}
\end{table}
\section{Conclusions and Future Work}
We proposed the concept of \textit{eventness} for audio events detection by utilizing a vision inspired CNN. Sharing similar characteristics with its vision based counterpart objectness in natural object detection, eventness can leverage components from natural object detection to detect audio events present in spectrograms.  We evaluated a Faster R-CNN adaptation for audio data in a qualitative experiment and a quantitative experiment. The results showed that the proposed eventness model detected audio events in spectrogram images comparable with the baseline model. Moreover, the eventness model is more robust to classifying infrequency events. Qualitative results also showed that the eventness model can exploit both the temporal and spectral content of the audio events.  

The work in this paper is a proof-of-concept for the eventness model. In the future, there are several other modifications or research directions that will be explored.  First, the feature maps are generated from a VGG-16 CNN that was pre-trained on the ImageNet dataset meaning that they are tailored to natural image representations.  Recently, the AudioSet dataset \cite{gemmeke2017audio} has been released with millions of audio clips thereby allowing for a CNN to be trained on purely audio data and removing the natural image representations inherent in the feature maps.  Other models to generate feature maps will also be explored that are not limited to the VGG-16 model.  Second, the current RPN using all locations on the feature map to generate proposals. However, the eventness model itself can determine areas on the spectrogram with the highest importance for the classification task.  We therefore propose to use \textit{attention} based models \cite{jaderberg2015spatial} to automatically learn the proposals.  Finally, we would like to modify the model such that a true end-to-end representation can be learned, i.e., where only the raw audio files are used as input instead of spectrograms as input to the Faster R-CNN.
%
\bibliographystyle{IEEEbib}
\bibliography{eventness}

\end{document}